\documentclass{WileyMSP-template}
\usepackage{mathalpha} 
\usepackage{graphicx} 
\usepackage{physics}
\usepackage{subcaption} 
\begin{document}

\pagestyle{fancy}
\rhead{\includegraphics[width=2.5cm]{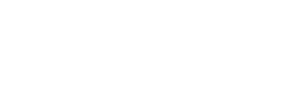}}

\title{Machine Learning-Based Optimization of Chiral Photonic Nano\-structures: Evolution- and Neural Network-Based Design}

\maketitle


\author{Oliver Mey*}
\author{Arash Rahimi-Iman*}

\begin{affiliations}
O. Mey\\
Fraunhofer IIS/EAS, Fraunhofer Institute for Integrated Circuits,\\
Division Engineering of Adaptive Systems, Dresden\\
Münchner Str. 16, 01187 Dresden, Germany\\~\\

A. Rahimi-Iman\\
1. Physikalisches Institut,\\
Justus-Liebig-Universität Gießen,\\
Gießen, 35390, Germany\\~\\

*correspondence: oliver.mey@eas.iis.fraunhofer.de, arash.rahimi-iman@physik.jlug.de
\end{affiliations}



\begin{abstract}
Chiral photonics opens new pathways to manipulate light--matter interactions and tailor the optical response of metasurfaces and -materials by nanostructuring nontrivial patterns. Chirality of matter, such as that of molecules, and light, which in the simplest case is given by the handedness of circular polarization, have attracted much attention for applications in chemistry, nanophotonics and optical information processing. We report the design of chiral photonic structures using two machine learning methods, the evolutionary algorithm and neural network approach, for rapid and efficient optimization of optical properties for dielectric metasurfaces. The design recipes obtained for visible light in the range of transition-metal dichalcogenide exciton resonances show a frequency-dependent modification in the reflected light's degree of circular polarization, that is represented by the difference between left- and right-circularly polarized intensity. Our results suggest the facile fabrication and characterization of optical nanopatterned reflectors for chirality-sensitive light--matter coupling scenarios employing tungsten disulfide as possible active material with features such as valley Hall effect and optical valley coherence. 
\end{abstract}

\section{Introduction}
Nanophotonic structures and optical elements with nontrivial properties and special functionalities have gained strong attention due to their application potentials in imaging, biomedicine, optoelectronics and quantum photonics. \\
Controlling light-matter interactions on sub-wavelength lengthscales requires advanced techniques regarding modelling, production and characterization --- typically addressed in the nanoworld due to the wavelength of visible or higher-energy light.  \\
New concepts for the design of structures and materials, as well as for the prediction of optical properties (also of light--matter interactions) have recently seen a strong demand. Methods based on computational optimization techniques (such as the adjoint method and inverse design) \cite{hughes2018}, machine learning \cite{kudyshev2020, pilozzi2018, liu2018, long2019} and deep learning \cite{yao2019, sajedan2019, hegde2020, nadell2019, chen2019, ma2018} (in some cases utilizing techniques such as deep reinforcement learning \cite{mnih2018}) have been introduced to tackle fundamental difficulties in the modelling and design of features, such as with regard to optical metasurfaces \cite{liu2018b, chen2019, nadell2019, kudyshev2020b}, nonlinear optics \cite{hughes2018}, topological photonics \cite{pilozzi2018, long2019} or chiral photonics \cite{ma2018}. \\
Furthermore, deep learning can significantly speed-up time-consuming optics design and the optimization process \cite{hedge2019} by assisting in finding an optics design starting point \cite{cote2021, cote2019, yang2019} and providing a potential alternative for the efficient inverse optical design \cite{zhang2020}. \\
While chirality is a fundamental structural property that tells whether or not a structure can be superimposed on its mirror image, the existence of such property gained attraction for applications. Chirality is understood to play an important role in chemical reactions of certain molecules \cite{bayku2019}, characteristic excitation pathways driven by circularly-polarized light fields \cite{kuznet2021, cao2012, mak2012} and in the manipulation as well as control of electro-magnetic radiation through tailored micro- and nanostructures (see literature examples below), which are commonly realized as metamaterials.

A simple and easy-to-imagine form of chiral light is given by left or right circularly polarized (short, LCP or RCP) light. Usually, optical interfaces or media have an impact on the reflected intensity as well as phase, by reflection and transmission or propagation, respectively, of polarized light.

Recently, studies involving systems with polarization-sensitive light--matter interactions, such as with chiral molecules \cite{yoo2015, venkata2017, stahl2020, gaul2020} or some of the popular valleytronic 2D materials \cite{mak2018, schaibley2016}, increased the demand for chiral photonics. Different concepts were under investigation in the past decade to address chirality of light, typically making use of chiral optical metamaterials of dielectric or metallic character that rely on nanostructuring materials \cite{collins2017, wang2016a}. Some designs make use of helix nanostructures for chirality at optical frequencies \cite{passaseo2017}, others target plasmonic nanophotonics \cite{hentschel2017} or 3D metaphotonic nanostructures \cite{qiu2018}. Also, for chiral light--matter interactions, synthetic chiral light fields were proposed \cite{ayuso2019}. Another aspect followed in the chiral optics domain is that of handedness of reflected light from chiral mirrors \cite{plum2015}, where it is discriminated between handedness preserving or nonpreserving reflection.

(Metallic) split ring resonators have been commonly developed to selectively absorb circularly polarized light \cite{plum2015, plum2016, jing2017, wang2016c, ye2017} and offer the possibility of a tight field confinement. These strong polarization-selective effects come with the disadvantage of Ohmic (i.e. charge carrier resistance dependent) losses, as incident fields cause oscillating electron clouds that dissipate energy through heat-generating scattering processes. Moreover, in nanostructured metamaterials, truncating the magnetic field additionally causes conversion of stored magnetic energy from the irradiated field into kinetic energy of electrons which causes a loss termed the Landau damping \cite{khurgin2015}. Dielectric chiral structures on the other hand bear the advantage that optical chirality can be addressed through selective reflection and transmission, while absorption may only play a minor role for such dielectric metasurfaces. They provide the opportunity to modify, as appropriate for a given application, reflection, transmission \cite{wu2014} and holographic \cite{khorasan2016} behavior and benefit from a negligible heat generation due to the low absorption by the metamaterial. Nonetheless, metallic metamaterials can be appealing for a different range of applications, e.g. for active chiral plasmonics that were reported with metal antenna metasurfaces \cite{yin2015}, whereas dielectric counterparts can excel via their passive nature of beam manipulation.

Previous works and theoretical considerations suggested that with (state-of-the-art/available) dielectric chiral photonics, one can---as envisioned for many years---design such optics with chiral reflectivity or transmittivity \cite{mai2019, semnani2019, zhu2018}.

Designing a suitable chiral structure with a set of desired (optical and geometrical) features is a complex problem. Even though certain requirements concerning symmetry are understood which enable one to assume the chiral nature of a structure \cite{wang2016c,plum2015}, there is a lack in our ability to make predictions owing to the complex, little known principles behind the achievement of a certain chiral behavior. A guideline or tool to solve this task of designing the chiral response of metamaterials and the behavioral changes upon structural transformations would be very helpful, and algorithmic approaches seem very reasonable.

\begin{figure}
\includegraphics[width=1.0\textwidth]{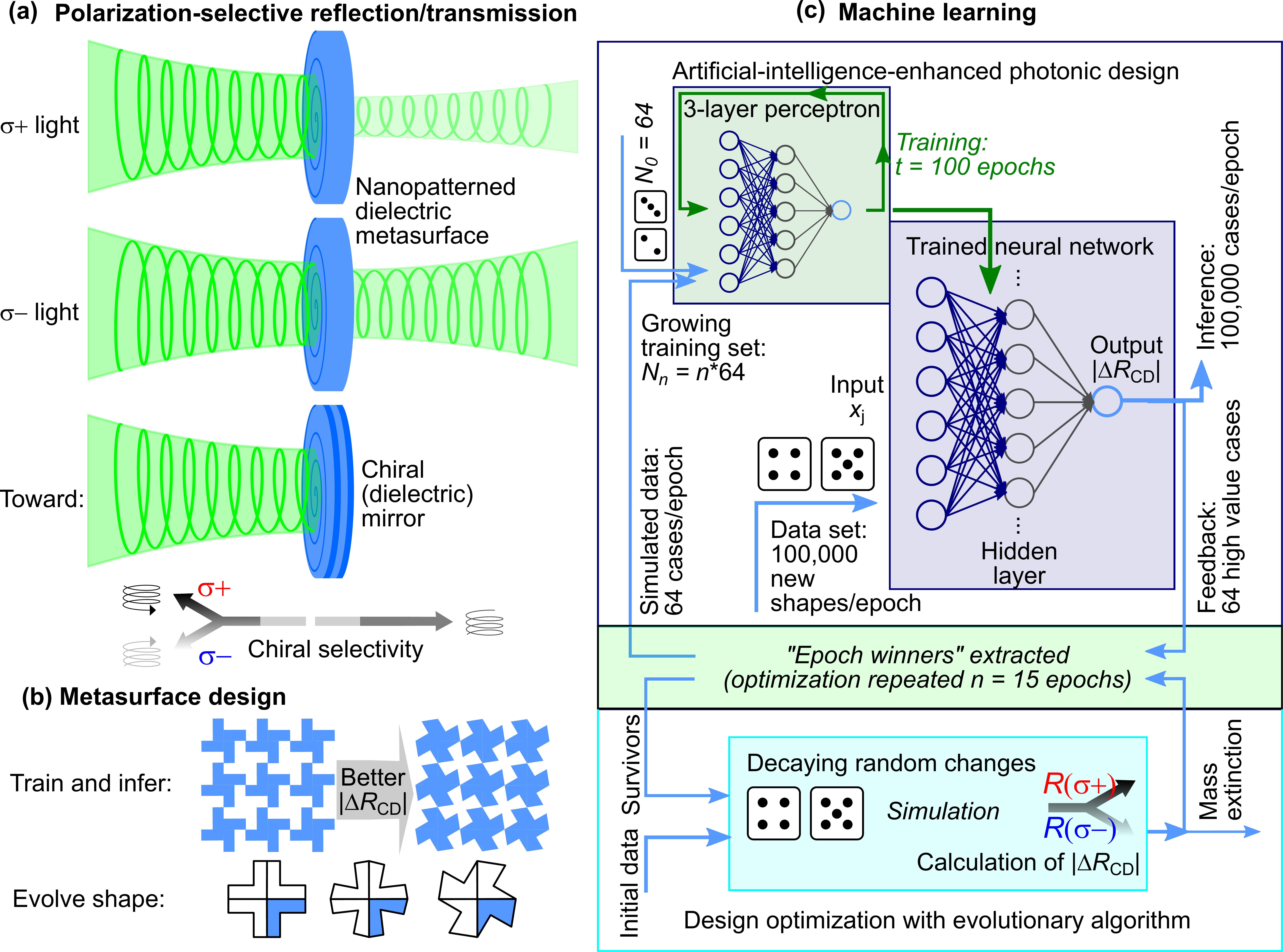}
\caption{Design goal and computer-assisted optimization concept. (a) Sketch of a polarization-dependent optical structure for chiral reflectivity/transmittance. Opposite circular polarizations of light are shown to experience different behavior of a chiral photonic structure. Toward the realization of chiral mirrors, optimized metasurface designs (b) are sought, here employing chiral periodic patterns. (c) Diagrammatic representation of two optimization schemes based on a perceptron-type neural network (top) and an evolutionary algorithm (bottom). Based on selective feedback (reinforcement learning) with epoch winners and a mass-extinction event for generated shapes after each optimization step, respectively, winner structures are extracted from the machine learning output.}
\label{fig_intro}
\end{figure}

Here, we present a methodology how to optimize chirality using an algorithmic optimization, as schematically represented in Figure \ref{fig_intro}. While next to our study related works have been discussing deep-learning architectures for on-demand design of chiral metamaterials, such as in Ref. \cite{ma2018}, this field is still in its infancy. At the design case of a dielectric interface, we compare two approaches to predict high chiral reflectivity, which is the intensity difference between LCP and RCP light: An evolutionary algorithm and a neural network based optimization. Our structure's spectral properties are adjusted to the optical spectrum of tungsten disulfide, a popular transition-metal dichalcogenide (TMDC) material for optoelectronic and nanophotonic applications. The resonance at the valley-polarized A-exciton emission is addressed by our design for future polarization-sensitive micro- and nanophotonics with TMDCs. These concepts can also benefit organic chiral materials \cite{venkata2017, yoo2015}, e.g. for biophysical applications. They promise tailored optical valleytronic behavior for light harvesting, storage and emission schemes and selective light--matter coupling experiments.

\section{Methods}
The process utilized in this work comprises alternating simulation and optimization steps. At first, shape geometries are proposed and their dielectric behavior is simulated. Using the simulations, their circular dichroism (CD) is evaluated. In a second step, an optimization algorithm is applied to find geometry proposals with a stronger CD in the next iteration. Next, we begin with an overview on how and under which constraints simulations were conducted (Section \ref{methods_simulation}). Following that, the two optimization algorithms are explained in detail, which are an evolutionary algorithm (Section \ref{methods_evolutionary}) and a neural network-based optimization (Section \ref{methods_nn}).
\subsection{Simulation of dielectric structures}
\label{methods_simulation}
To simulate the dielectric behavior of nanostructured photonic elements, the software S$^4$ \cite{liu2012} was employed. This software implements the Fourier modal method \cite{li1997} and thereby enables the calculation of reflectivity and transmission spectra of layered dielectric materials, which feature an in-plane periodic structure. In S$^4$, the two-dimensional periodic structures of the complex relative permittivity are approximated by Fourier series expansions of its geometry. This implies that calculations get more accurate, the more terms of the Fourier series are used to represent the geometry. On the other hand, the required computation time scales with the third potency of the number of used Fourier orders \cite{liu2012}. A trade-off therefore needs to be found which allows for a reasonable computation time but still enables accurate calculations. In this work, a number of $n_F=200$ Fourier orders was used for calculation. This is twice the number of Fourier coefficients which was suggested by the authors of S$^4$ for use with simpler 2D structures like arrays of circular holes. The higher number of Fourier orders was chosen due to the higher geometric complexity of the simulated structures due to the geometric constraints explained in the following.\\
To create photonic structures with circular polarization selectivity, certain geometric conditions were derived in \cite{wang2016b}. For circular polarization selective mirrors, no mirror symmetry is allowed at all. To have handedness preserving mirrors, no $n_{rot}$-fold rotational symmetry with $n_{rot}>2$ is allowed. To have handedness reversing mirrors, all kinds of rotational symmetry are allowed. Here we chose to optimize towards a handedness reversing mirror, and therefore we aimed for structures, with a rotational symmetry with $n_{rot}>2$. To minimize the number of parameters to optimize, we selected $n_{rot}$ to equal 4. Doing this, only parameters in the first quadrant of the unit cell need to be optimized by the algorithms since the geometry in all other quadrants follow by exploiting the selected rotational symmetry. The geo\-metries were further constrained to have 3 corners. Since every corner has x- and y- coordinates, there were 6 parameters $x_j$ with $j\in\left\lbrace 1,\ldots, 6 \right\rbrace $ to be optimized. By randomly setting the coordinates of the corners of the chiral elements, intersections between lines connecting two points can appear and need to be prevented. In addition, it is reasonable to determine a minimum angle $\alpha_\mathrm{min}$, which two connecting lines may enclose at a corner. The smaller $\alpha_\mathrm{min}$ is, the higher requirements are placed on the minimum structure sizes, when the resulting patterns are to be produced through suitable nanostructuring techniques. Before simulations are conducted, each geometry proposal is checked to not include intersections and that $\alpha_\mathrm{min}>45°$. If these conditions are not met, new randomly generated geometries are created until there is one which meets the formulated requirements.\\
It is expected that a strong CD can be achieved by utilizing large refractive index contrasts. On the other hand, the CD should be accomplished without utilizing absorption effects but by only increasing the differences in terms of reflectivity and transmittivity between LCP and RCP light. A material which exhibits a large refractive index and nearly no absorption over a large part of the visible range is Gallium phosphide (GaP). It has a refractive index of 3.34 for light at 615\,nm \cite{aspnes1983}, which is the photoluminescence wavelength of WS$_2$. A structured GaP substrate was therefore chosen as material for the investigation. The corners in all 4 quadrants build up a polygon, which is cut out of this GaP substrate. The parts which are cut out are modeled with a refractive index of $1.0$ (corresponding to air/ vacuum).\\
The whole simulation was kept unitless. This means that the frequency $f$ and wavelength $\lambda$ of the incident light was set to 1. The lattice constant $l$ of the periodic structure was set to $l=2\cdot \lambda$, and the depth of the structure to $0.25\cdot \lambda$. For the case $\lambda=615\,$nm, this would correspond to a lattice constant of 1230\,nm and a structure depth of 153.5\,nm, which is in the range of currently available typical (lithographic) structuring methods such as focused ion-beam milling or photolithography.\\
To evaluate the CD, the integral
\begin{equation}
\Delta R_{CD}=\int_{f_{\mathrm{low}}}^{f_{\mathrm{high}}} \left( R_{\mathrm{LCP}}(f)-R_{\mathrm{RCP}}(f) \right) \dd{f}
\end{equation}
was calculated. $R_{\mathrm{RCP}}$ and $R_{\mathrm{LCP}}$ are the reflectivities for LCP and RCP light, respectively. The integral was evaluated in the frequency range between $f_{\mathrm{low}}=0.95$ and $f_{\mathrm{high}}=1.05$ with a stepsize of 0.002. Due to the small evaluated frequency range, the refractive indices $n_{\mathrm{GaP}}$ and $n_{\mathrm{air}}$ were approximated as static with $n_{\mathrm{GaP}}=3.34$ and $n_{\mathrm{air}}=1.0$.
\subsection{Evolutionary Optimization}
\label{methods_evolutionary}
As a first approach for the optimization, an algorithm was constructed which is based on the principle of evolution. At the beginning, an initial population of $N_{0}=64$ randomly generated geometries is simulated and the value of the target variable $\Delta R_{CD}$ is calculated for each of them. The $N_{\mathrm{best}}=4$ patterns, which show the highest value for $|\Delta R_{CD}|$, are selected to survive a mass extinction event. For each following epoch $i$ with $i\in\left\lbrace 1,\ldots, n_{\mathrm{epochs}} \right\rbrace $ and the number of epochs $n_{\mathrm{epochs}}$, the $N_{\mathrm{best}}$ shapes which had the highest value of $\Delta R_{CD}$ throughout the whole optimization process were extracted and $N_i$ new patterns were created from those best shapes by applying randomly generated geometry changes to them. For the epoch $i$, the new coordinates $x_{j,i}$ were derived using following equation:
\begin{equation}
x_{j,i}=x_{j,i-1} + \frac{l}{\eta_{0}^{i}}\cdot x_{\Delta j,i}.
\label{geometry_evolution_eq}
\end{equation}
The parameter $\eta_0$ thereby describes the decay rate for an exponentially decreasing amount of geometry variation and $x_{\Delta j,i}$ are randomly generated values drawn from a uniform distribution $\mathcal{U}(a=-l/2,\allowbreak b=+l/2 )$ with the lower and upper boundaries of the distribution $a$ and $b$. $x_{j,i-1}$ describe the $j$ coordinate values from the shape whose geometry is to be changed. The extent, to which the geometries are varied, thereby decreases exponentially throughout the evolution process. Using (\ref{geometry_evolution_eq}) will in many cases lead to coordinates which lie outside the unit cell or violate the geometry conditions regarding the minimum allowed angles and the necessecary non-intersection of the borders of the resulting shapes. For each proposed new shape, all geometry checks therefore have to be repeated. Whenever a certain condition is not met, new random values are drawn and the checks are repeated until $N_{i}$ new shapes are found which fulfill all conditions. For each of them, the corresponding S$^4$ simulation is executed as described in Section \ref{methods_simulation} in order to obtain their respective values of $\Delta R_{CD}$.\\
In our experiments, we used values of $n_{\mathrm{epochs}}=15$ and $\eta_{0}=1.45$ and simulated $N_0 = N_i =64$ different structures in each epoch.
\subsection{Neural Network-based Optimization}
\label{methods_nn}
The second optimization approach utilized a reinforcement learning methodology. As for the evolutionary optimization approach, in the first epoch, the value of $\Delta R_{CD}$ is simulated for $N_{0}=64$ randomly generated geometries. Afterwards, a neural network is trained to predict the CD integral of each structure $\Delta R_{CD}$ by having the standardized 6 coordinate values $x_j$ of each shape as input. The neural network used is a 3-layer perceptron with 16 units in its hidden layer and a rectifier activation function (ReLU) in its hidden layer. This comparatively small neural network architecture was chosen to prevent overfitting, as only a low number of input and output variables had to be processed. Also, it was aspired to have a low computational effort during inference. The neural network was trained for $n_{\mathrm{epochs,NN}}=100$ epochs by minimizing the mean squared error of the predicted $\Delta R_{CD}$ values using the backpropagation algorithm. To get samples for the next iteration of the optimization algorithm, the $\Delta R_{CD}$ values of $N_{\mathrm{inference}}=100{,}000$ randomly generated shapes were predicted by the trained neural network. As with the proposed shapes in Section \ref{methods_evolutionary}, the randomly generated shapes thereby had to be non-intersecting and abide by the aforementioned minimum angle criterion. From the $N_{\mathrm{inference}}$ shapes, the $N_{0}=64$ shapes with the highest predicted value of $|\Delta R_{CD}|$ were extracted and an S$^4$ simulation as described in Section \ref{methods_simulation} was conducted to determine their level of CD according to the simulation. For every following epoch of the optimization, the already pretrained neural network is retrained with all available simulated pairs of coordinates and corresponding $\Delta R_{CD}$ values, which means that the number of samples available for the training of the neural network scales linearly with the number of iterations already conducted. The neural network can thereby learn to correct its predictions with the highest predicted CD values from the previous iterations.\\
Similar to the approach described in Section \ref{methods_evolutionary}, the optimization was conducted for $n_{\mathrm{epochs}}=15$ epochs. For an example of the obtained progression of optimized shapes over a number of epochs, see the Supporting Information section.
\section{Results}
With both of the two above described algorithms, an optimization was conducted to propose suitable geometrical shapes for an in-plane periodic structured pattern on the surface of a dielectric material, here GaP, for which a strong CD is predicted in the spectral vicinity of a design frequency. Next, we compare characteristics of the optimization process obtained thereby.
\subsection{Evolutionary Optimization}
\label{results_evolutionary}
\begin{figure}
\includegraphics[width=1.0\textwidth]{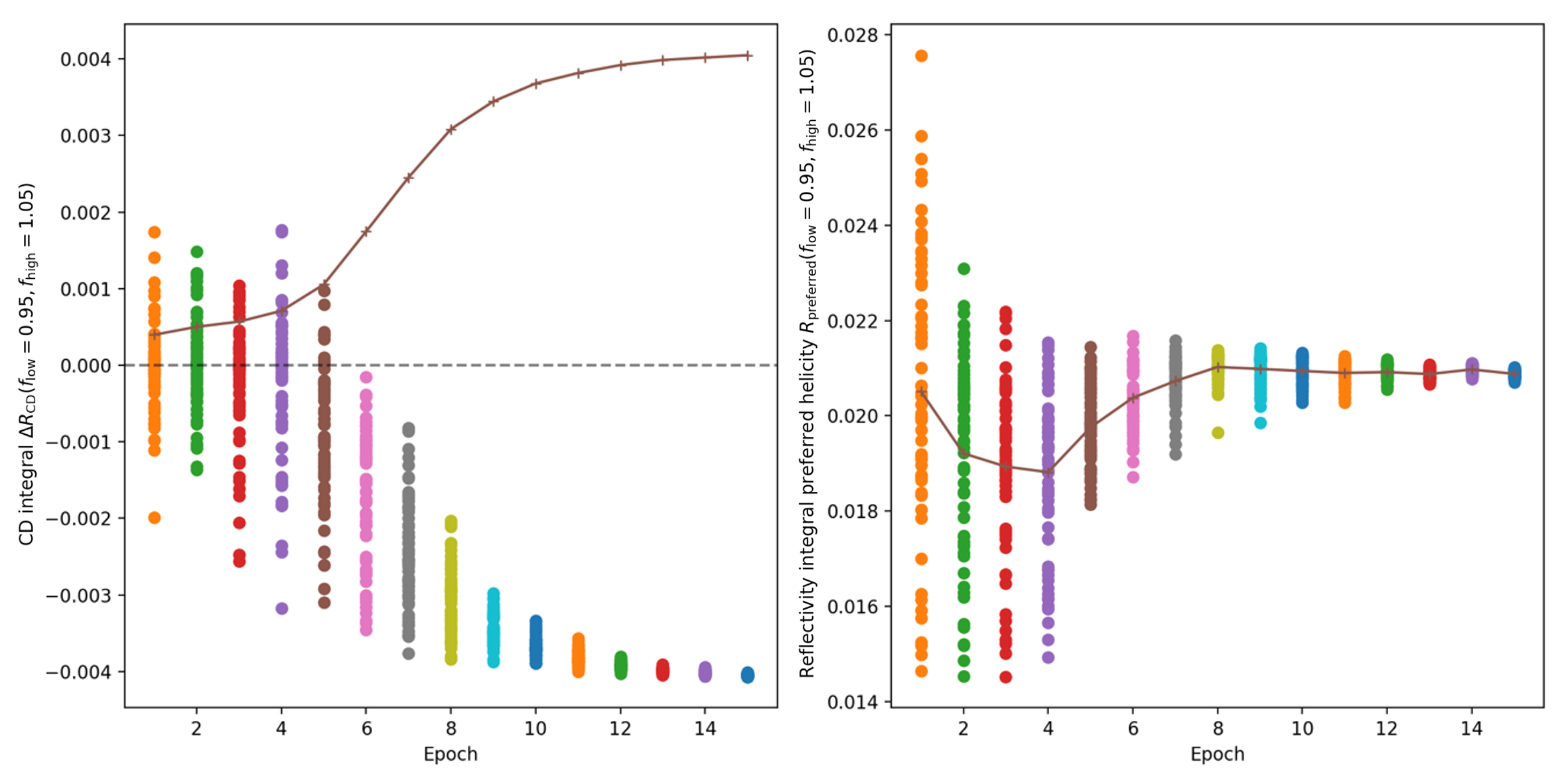}
\caption{Results of the evolutionary optimization. Left: Progression of the chiral dichroism (CD) integral $\Delta R_{CD}$, which represents the difference between LCP-RCP reflectivities spectrally-integrated around a target frequency. Here, the negative maximum corresponds to an average of around 4 percent points. Right: Progression of spectrally integrated reflectivity around the same target frequency for the simulated spectrum showing the calculated absolute value of the preferred polarization. Here, the convergence goes to an average reflectivity of 21 \%. In both cases, dots represent individual values obtained by simulation, the plotted line represents the mean value of the individual absolute values per epoch.}
\label{evolution_results_figure}
\end{figure}

The left part of Figure \ref{evolution_results_figure} depicts the progression of the simulated value of $\Delta R_{CD}$ throughout the optimization for the case of the evolutionary optimization.
There, each dot represents one conducted simulation. In the first epoch, where all shapes were selected without prior knowledge, the resulting $\Delta R_{CD}$-values are centered around zero with a certain variance. The strongest CD achieved in the first epoch had a $\Delta R_{CD}$-value of around $-0.002$. The negative sign means that RCP light is reflected with a higher reflectivity compared to LCP light. With increasing number of epochs finished, the value of $\Delta R_{CD}$ progressed towards higher magnitudes, while its variance decreased. While the lower variance for higher epoch numbers can be explained by the decreasing amount of geometry variation controlled by the decay parameter $\eta_0$, the progression of the center of the $\Delta R_{CD}$-values shows the effectivity of the proposed method: The shapes simulated in the last epoch have a CD twice as strong as the strongest CD in the first epoch. While during the first 5 epochs both RCP- and LCP- preferring geometries are simulated, there are only RCP-preferring shapes in epoch 6 and thereafter. However, this convergence towards RCP-preferring geometries happened only by chance, reversing the chirality of a certain shape would lead to the same absolute value of $\Delta R_{CD}$ but with opposite sign. Besides the reflectivity difference between RCP and LCP light, also the respective reflectivity of the preferred circular polarization was investigated. That is, the integral
\begin{equation}
R_{\mathrm{preferred}} = \max \left( \int_{f_{\mathrm{low}}}^{f_{\mathrm{high}}} R_{\mathrm{RCP}}(f) \dd{f}, \int_{f_{\mathrm{low}}}^{f_{\mathrm{high}}} R_{\mathrm{LCP}}(f) \dd{f} \right)
\end{equation} was calculated and plotted in the right part of Figure \ref{evolution_results_figure}. As with the reflectivity difference, at the beginning of the optimization there is a large variance which decreases with increasing duration of the optimization due to the decreasing amount of geometry variation. The population-per-epoch mean value of $R_{\mathrm{preferred}}$ is, however, nearly the same at the beginning and the end of the optimization. This means that the optimization of the shapes led to a constant reflection for RCP light and a higher transmission for LCP light.
\subsection{Neural Network-based Optimization}
\label{results_nn}
\begin{figure}
\includegraphics[width=1.0\textwidth]{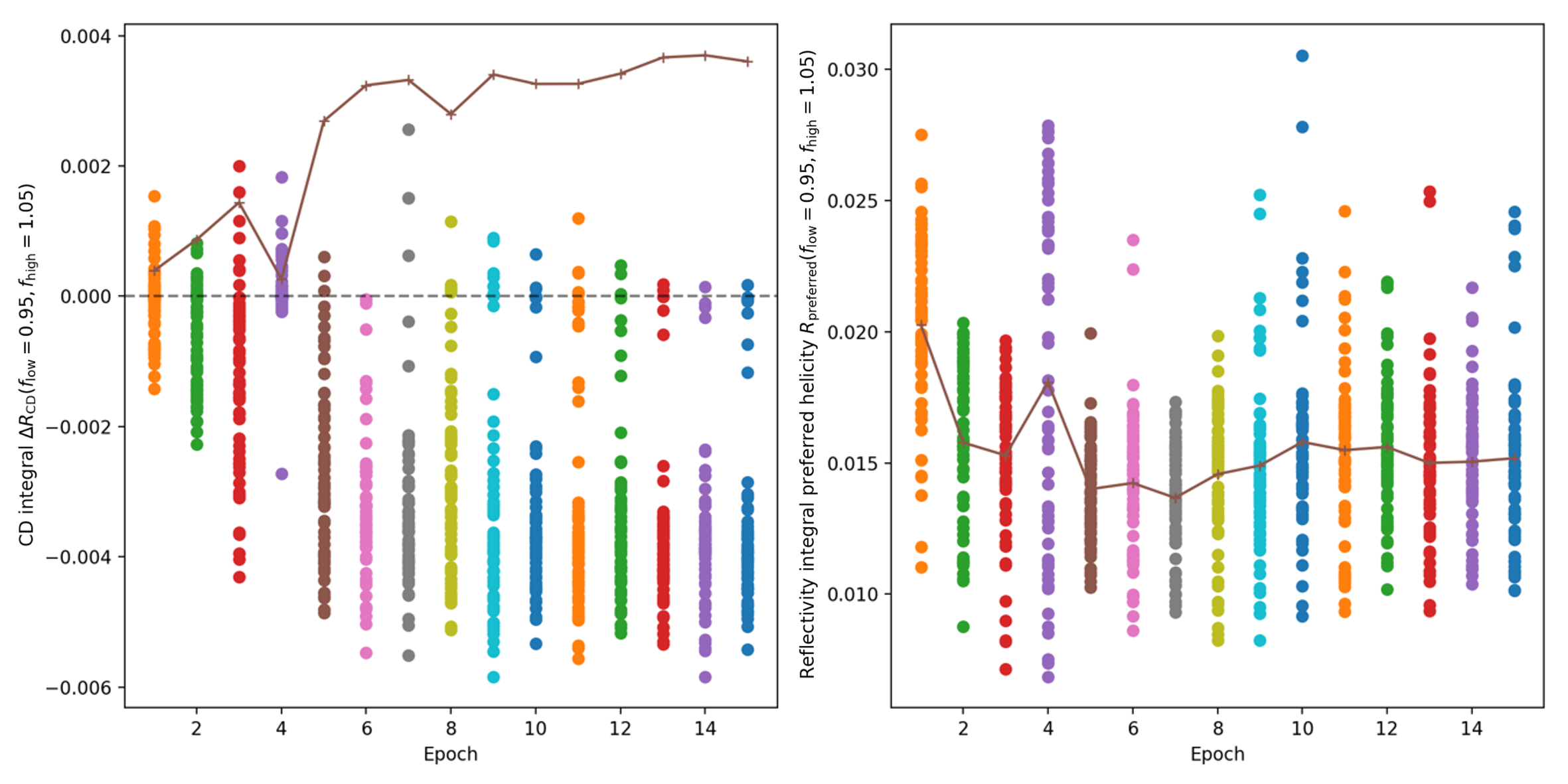}
\caption{Results of the neural network-based optimization. Left: Progression of the chiral dichroism (CD) integral $\Delta R_{CD}$, which represents the difference between LCP-RCP reflectivities spectrally-integrated around a target frequency. Here, the negative maximum corresponds to an average of around 6 percent points. Right: Progression of spectrally integrated reflectivity around the same target frequency for the simulated spectrum showing the calculated absolute value of the preferred polarization. Here, the convergence goes to an average reflectivity of 15 \%. The representation is similar to Figure \ref{evolution_results_figure}.}
\label{nn_results_figure}
\end{figure}

The same types of diagrams presented before in Section \ref{results_evolutionary} but now applied to the neural network-based optimization are given in Figure \ref{nn_results_figure}.
 Starting with the evaluation of the reflectivity difference depicted in the left panel, it directly stands out that the variance of the $\Delta R_{CD}$ values of the individual simulations does not decrease but even increase during the course of the optimization. The fundamental difference with the neural network-based approach is that still in the last epoch of the optimization, a huge variety of different shapes can be proposed as promising by the algorithm. This is in contrast to the working mechanism of the evolutionary algorithm, where the already best performing patterns are modified less the further the optimization proceeds. Since the multilayer perceptron utilized is apparently not able to perfectly learn the relationship between input coordinates and $\Delta R_{CD}$ outputs, there are also shapes proposed in each epoch which do not have a strong CD. On the other hand, the neural network-based optimization is able to propose shapes with a much stronger CD than the evolutionary optimization algorithm. The best result from this optimization has an $\Delta R_{CD}$ value of nearly 0.006, which is about 50\,\% higher than the best result from the evolutionary optimization. This value, however, is already achieved after nine optimization epochs. Also, the evaluation of the mean of the absolute values of $\Delta R_{CD}$ shows that there occurs a certain saturation after about 7--9 epochs.\\
An examination of the reflectivity of the preferred light orientation illustrates again the higher variability of the individual results of each optimization epoch. However, for this type of optimization the overall reflectivity seems to decrease. Apparently, it is easier for the algorithm to find shapes with a strong CD at lower reflectivity levels, or in other terms, obtaining a stronger dichroism seems to be linked to a reduction in reflectivity for the tailored interface, as is also later shown in spectra corresponding to the resulting optimized shape.
\section{Discussion}
\begin{figure}[h!]
\captionsetup[subfigure]{justification=centering}
\begin{subfigure}[t]{0.395\textwidth}
\includegraphics[width=\linewidth]{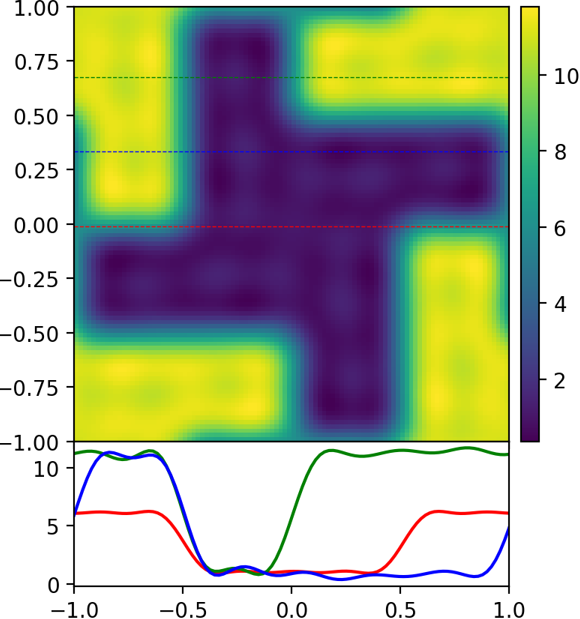}
\caption{}
\label{windmill_shape}
\end{subfigure}
\hspace*{\fill}   
\begin{subfigure}[t]{0.59\textwidth}
\includegraphics[width=\linewidth]{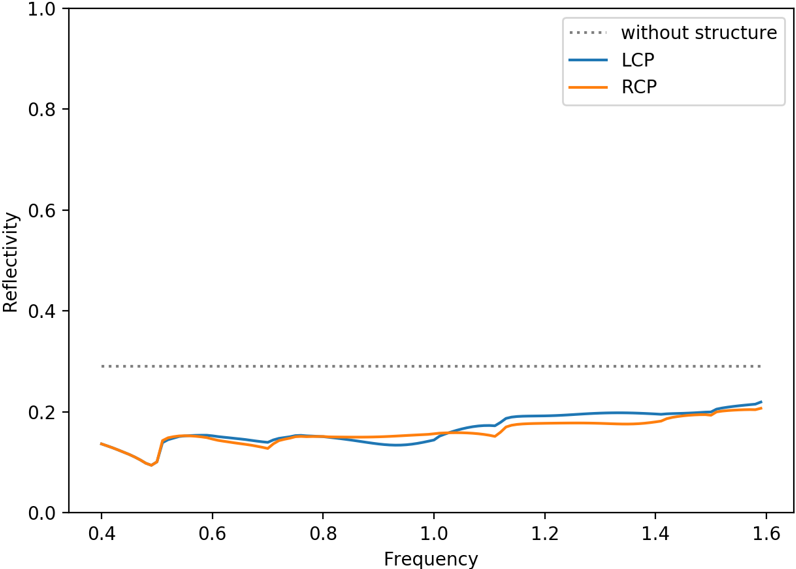}
\begin{picture}(0,0)
\put(30,192){\includegraphics[height=1.5cm]{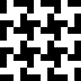}}
\end{picture}
\caption{}
\label{windmill_spectrum}
\end{subfigure}
\\
\begin{subfigure}[t]{0.395\textwidth}
\includegraphics[width=\linewidth]{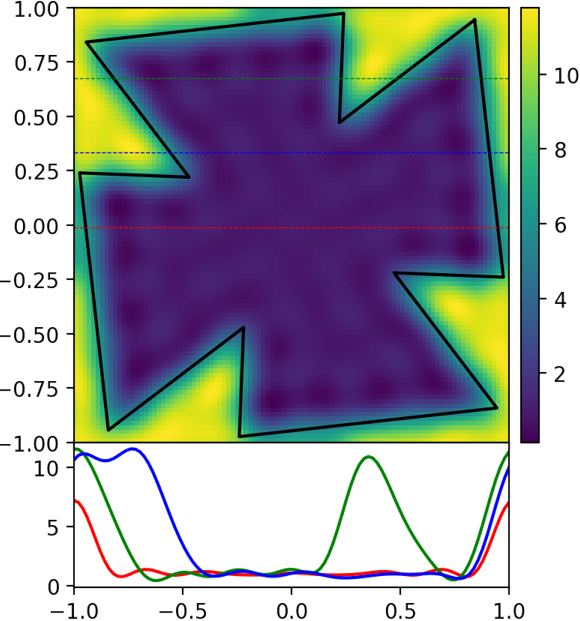}
\caption{}
\label{nn_winner_shape}
\end{subfigure}
\hspace*{\fill}   
\begin{subfigure}[t]{0.59\textwidth}
\includegraphics[width=\linewidth]{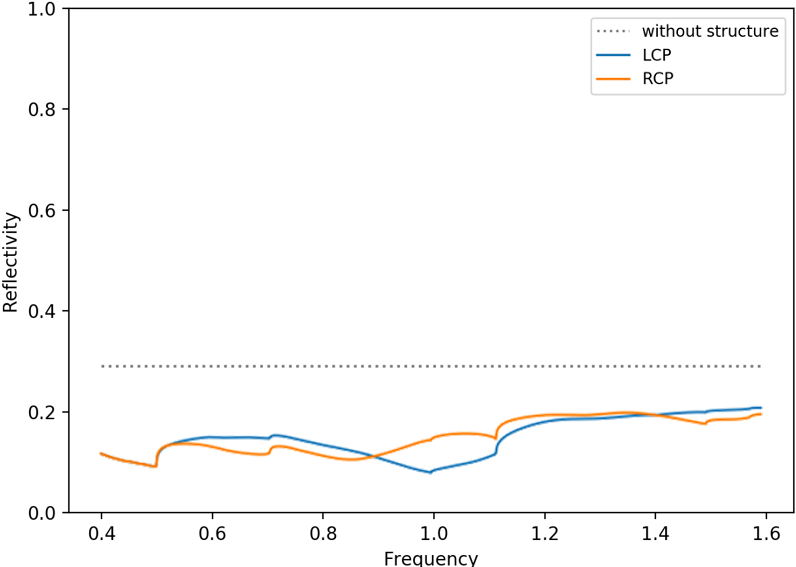}
\begin{picture}(0,0)
\put(30,192){\includegraphics[height=1.5cm]{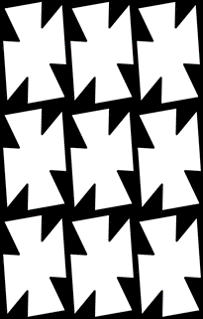}}
\end{picture}
\caption{}
\label{nn_winner_spectrum}
\end{subfigure}
\caption{Dielectric landscape of one unit cell of a hand-designed shape (a) and the shape with the highest CD from the simulation (c). The curves below are cross sections with the respective positions indicated as dashed lines in the images above. The respective reflectivity spectrums for LCP and RCP light are given in (b) and (d). The black dashed line indicates the reflectivity of an unstructured air-GaP surface. Insets show the visualize the resulting pattern of 3x3 lattice cells.}
\end{figure}

To evaluate the effectiveness of the proposed optimization algorithms, the best result from the neural network-based optimization was compared with a hand-designed chiral pattern. First, we introduce a reference structure for a periodic surface pattern with expected chiral optical properties. Its relative permittivity map of one unit cell is depicted in Figure \ref{windmill_shape}. Due to its visual appearance, we refer to it as \textit{windmill shape}. This straight-forwardly hand-designed intuitive geometry has only orthogonal corners and its arms have a width of one half of the wavelength $\lambda_0$ in air. The area where the planar GaP substrate is structured with $n=n_\mathrm{air}$ thereby has exactly the same shape as the unstructured areas with $n=n_\mathrm{GaP}$. Note that this choice is made for the sake of simplicity and facile sample production; it bears no further medium-specific wavelength-related adjustments of feature sizes and can conveniently serve as reference structure (both for theory work and practical examination). The resulting reflectivity spectrum of the corresponding dielectric metasurface structured with the windmill shape is shown in Figure \ref{windmill_spectrum}. Indeed, there is a certain CD with an overall reduced reflectivity.\\
In contrast, the dielectric landscape of one unit cell of the structure, which exhibited the strongest CD throughout the optimization procedures, is visualized in Figure \ref{nn_winner_shape}. This shape will be referred to as \textit{winner shape}. As can be seen from its false-color map, the region for this shape with $n=n_\mathrm{air}$ (dark color) fills a larger area of the unit cell than those with $n=n_\mathrm{GaP}$ (bright color). This is reasonable, since the wavelength inside the dielectric medium is inversely proportional to the refractive index of that material. The incident light's electromagnetic field therefore oscillates on smaller length scales inside GaP compared to the case in air. The reflectivity spectrum corresponding to the dielectric metasurface based on such pattern structure is depicted in Figure \ref{nn_winner_spectrum}. Compared to the windmill shape, there is a much stronger CD. As with the windmill shape, also for this case, the overall reflectivity is reduced. 
\\Comparing both optimization approaches, we evidenced that the structure obtained by utilizing the neural network-based algorithm exhibited a CD about 50\,\% stronger than for the design optimization suggested by the evolutionary algorithm with the same number of epochs. While both algorithms are able to find patterns with an increased selectivity regarding differently circularly polarized light, further investigations on these or extended optimization approaches remain promising. On the one hand, the usage of more sophisticated evolutionary algorithms like CMA-ES \cite{hansen2001, hansen2003} could possibly reverse the ranking between the evolutionary and the neural network-based approach. This may be particularly anticipated for specific optimization tasks where good initial conditions can be predicted, such as the windmill case or related chiral structures, which are based on prerequisites for their chiral behavior. On the other hand, further experiments on the neural network-based optimization could lead to improved results for this approach, as well. For example, a larger neural network --- at the cost of computational expense as well as possible overfitting --- could be able to learn more complex relationships between the coordinate values and the $\Delta R_{CD}$ output. This may be meaningful both for a small number of coordinate points as used here and for the case when more coordinate values from more complex geometries are fed into the network. Converting the definition of the geometries to a pixel-wise binary setting would enable the usage of convolutional neural networks. Further, the trained neural network could be investigated by calculating the Shapley values \cite{shapley2016, lundberg2017} of its predictions to find out which coordinate value influenced the output value most. Instead of randomly probing thousands of different geometries with the trained neural network, activation maximization \cite{erhan2009, zeiler2014} could be used to find out the coordinate values, for which the highest CD is predicted. Besides the transfer to a pixel-wise binary definition of the geometries, also the introduction of more than 3 corners inside one quadrant or the use of other design elements inside the unit cell, such as circular, elliptical or dash-like holes (with their central coordinates, dimensions and even with additional rotation degree of freedom), would be feasible. Thereby, the optimization of shapes with higher complexity can be enabled. In addition, also the material depth of the metasurface structure, which was set to one quarter of the wavelength here, could be a parameter to be optimized.\\
One limitation of the results presented here is that the findings solely base on simulations which employ the Fourier modal method. The results thereby depend strongly on the number of employed Fourier coefficients $n_F$ used to get a representation of the inplane periodic structure. The higher $n_F$ is, the more accurate the geometry can be represented and the sharper the refractive index shapes and features are in the Fourier representation, but the higher the computational intensity becomes. On the other hand, real lithographically produced samples with the proposed structures will actually never have perfectly sharp structural edges and correspondingly no perfectly sharp transitions from one to another refractive index value. The seemingly approximated representation of the geometry by Fourier series can therefore be even more realistic compared to a perfect geometry representation, that means with very sharp transitions in any spatial direction. Nevertheless, experimental investigations as well as the employment of other simulation approaches like FDTD \cite{oskooi2010} are necessary to assess the predicted chirality results reported here and to jugde on the optimization capabilities in practical scenarios involving (nano)structuring. It is worth noting that since shapes with orthogonal corners are easier to produce lithographically, an investigation with the constraint of orthogonal corners could make the outcome even more relevant for actual devices.\\
Besides the optimization of the CD in (nano)structured dielectric substrates, the demonstrated methodology is transferrable to a wide range of design tasks. First of all, the capabilities of existing and typically hand-designed split-ring resonator structures for handedness preserving or reversing mirrors mentioned in the introduction could be further enhanced by employing the proposed algorithmic optimization. Further, devices like photonic crystals or various metamaterials could be designed with our methodology. Even though the optimization algorithms have not been fine-tuned in every aspect yet, this study provides the proof-of-principle and motivates further investigations in this direction. 


\section{Conclusion}
The great demand for a cost-efficient, rapid and effective optimization of photonic structure designs for various applications encouraged the employment of machine learning techniques to assist engineers and scientists seeking a specific (set of) optical feature(s). Moreover, it promises automation of design and characterization processes.

Here, we show a proof-of-principle examination on how to use two different machine learning techniques for the optimization of chiral photonic dielectric metamaterials, yielding a pronounced difference in the reflectivity (and correspondingly also transmittivity) for the two opposite circular polarizations. Therefore, we have combined electromagnetic simulations with both an evolutionary and a neural-network-based optimization algorithm. 

The comparison between the two optimization methods shows the strength of this concept by the example of a winner structure with improved spectral signatures with regard to optical chirality for an initial design frame. Thus, algorithmic methods qualify for rapid optimization of optical properties for metasurfaces and -materials towards the design of, for instance, chiral mirrors. Furthermore, they can be adjusted according to the desired optical and geometrical features of a system and also take into consideration production aspects. 

Ultimately, our study motivates further improvement and application of this technique in the domain of photonic engineering as well as metamaterial and nanotechnological designs with the aim to tailor optics designs or modify surface, interface and material properties as desired.


\medskip
\textbf{Supporting Information} \par 
Supporting Information accompanies this paper.

\medskip
\textbf{Acknowledgements} \par 
Financial support by the Deutsche Forschungsgemeinschaft (DFG: RA2841/12-1) is acknowledged.
The authors are grateful to their former team members F. Wall, K.A. Fedorova and L.M. Schneider, for fruitful discussions on learning algorithms, machine learning and 2D materials, respectively, as well as literature suggestions.\medskip

\textbf{Conflict of Interest} \par 
The authors declare no conflict of interest.\medskip

\textbf{Author contributions}\par 
ARI initiated the study and conceived the concepts for nanopatterned dielectric metastructures for chiral photonics with 2D materials and nanoparticles. OM performed the theoretical photonic analysis,\linebreak evolution- and neural network-based optimization, and simulated polarization-dependent spectral properties under the guidance of ARI. OM and ARI designed the structures, outlined the theoretical work, as well as optimization approaches, and evaluated the data. The results were summarized in a manuscript by both authors. 
\medskip

\textbf{Corresponding author}\par 
oliver.mey@eas.iis.fraunhofer.de, arash.rahimi-iman@physik.jlug.de


\begin{thebibliography}{00}
\bibitem{hughes2018} T. W. Hughes, M. Minkov, I. A. D. Williamson, S. Fan, \textit{ACS Photonics}. \textbf{2018} \textit{5}, 4781, DOI: 10.1021/acsphotonics.8b01522.
\bibitem{kudyshev2020} Z. A. Kudyshev, V. M. Shalaev, A. Boltasseva, \textit{ACS Photonics}. \textit{2021} \textit{8}, 34, DOI: 10.1021/acsphotonics.0c00960.
\bibitem{pilozzi2018} L. Pilozzi, F. A. Farrelly, G. Marcucci, C. Conti, \textit{Commun Phys}. \textbf{2018} \textit{1}, DOI: 10.1038/s42005-018-0058-8.
\bibitem{liu2018} D. Liu, Y. Tan, E. Khoram, Z. Yu, \textit{ACS Photonics}. \textbf{2018} \textit{5}, 1365, DOI: 10.1021/acsphotonics.7b01377.
\bibitem{long2019} Y. Long, J. Ren, Y. Li, H. Chen, \textit{Appl. Phys. Lett.} \textbf{2019} \textit{114}, 181105, DOI: 10.1063/1.5094838.
\bibitem{yao2019} K. Yao, R. Unni, Y. Zheng, \textit{Nanophotonics}. \textbf{2019} \textit{8}, 339, DOI: 10.1515/nanoph-2018-0183.
\bibitem{sajedan2019} I. Sajedian, H. Lee, J. Rho, \textit{Scientific reports}. \textbf{2019} \textit{9}, 10899, DOI: 10.1038/s41598-019-47154-z.
\bibitem{hegde2020} R. S. Hegde, \textit{Nanoscale Adv}. \textbf{2020} \textit{2}, 1007, DOI: 10.1039/C9NA00656G.
\bibitem{nadell2019} C. C. Nadell, B. Huang, J. M. Malof, W. J. Padilla, \textit{Optics express}. \textbf{2019} \textit{27}, 27523, DOI: 10.1364/OE.27.027523.
\bibitem{chen2019} Y. Chen, J. Zhu, Y. Xie, N. Feng, Q. H. Liu, \textit{Nanoscale}. \textbf{2019} \textit{11}, 9749, DOI: 10.1039/c9nr01315f.
\bibitem{ma2018} W. Ma, F. Cheng, Y. Liu, \textit{ACS nano}. \textbf{2018} \textit{12}, 6326, DOI: 10.1021/acsnano.8b03569.
\bibitem{mnih2018} V. Mnih, K. Kavukcuoglu, D. Silver, A. A. Rusu, J. Veness, M. G. Bellemare, A. Graves, M. Riedmiller, A. K. Fidjeland, G. Ostrovski, S. Petersen, C. Beattie, A. Sadik, I. Antonoglou, H. King, D. Kumaran, D. Wierstra, S. Legg, D. Hassabis, \textit{Nature}. \textbf{2015} \textit{518}, 529, DOI: 10.1038/nature14236.
\bibitem{liu2018b} Z. Liu, D. Zhu, S. P. Rodrigues, K.-T. Lee, W. Cai, \textit{Nano letters}. \textbf{2018} \textit{18}, 6570, DOI: 10.1021/acs.nanolett.8b03171.
\bibitem{kudyshev2020b} Z. A. Kudyshev, A. V. Kildishev, V. M. Shalaev, A. Boltasseva, \textit{Applied Physics Reviews}. \textbf{2020} \textit{7}, 21407, DOI: 10.1063/1.5134792.
\bibitem{hedge2019} R. S. Hegde, \textit{Opt. Eng.} \textbf{2019} \textit{58}, 1, DOI: 10.1117/1.OE.58.6.065103.
\bibitem{cote2021} G. Côté, J.-F. Lalonde, S. Thibault, \textit{Optics express}. \textbf{2021} \textit{29}, 3841, DOI: 10.1364/OE.401590.
\bibitem{cote2019} G. Côté, J.-F. Lalonde, S. Thibault, \textit{Optics express}. \textbf{2019} \textit{27}, 28279, DOI: 10.1364/OE.27.028279.
\bibitem{yang2019} T. Yang, D. Cheng, Y. Wang, \textit{Optics express}. \textbf{2019} \textit{27}, 17228, DOI: 10.1364/OE.27.017228.
\bibitem{zhang2020} S. Zhang, L. Bian, Y. Zhang, \textit{J. Opt}. \textbf{2020} \textit{22}, 105802, DOI: 10.1088/2040-8986/abb1ce.
\bibitem{bayku2019} D. Baykusheva, D. Zindel, V. Svoboda, E. Bommeli, M. Ochsner, A. Tehlar, H. J. Wörner, \textit{Proceedings of the National Academy of Sciences of the United States of America}. \textbf{2019} \textit{116}, 23923, DOI: 10.1073/pnas.1907189116.
\bibitem{kuznet2021} V. Kuznetsova, Y. Gromova, M. Martinez-Carmona, F. Purcell-Milton, E. Ushakova, S. Cherevkov, V. Maslov, Y. K. Gun’ko, \textit{Nanophotonics}. \textbf{2021} \textit{10}, 797, DOI: 10.1515/nanoph-2020-0473.
\bibitem{cao2012} T. Cao, G. Wang, W. Han, H. Ye, C. Zhu, J. Shi, Q. Niu, P. Tan, E. Wang, B. Liu, J. Feng, \textit{Nature communications}. \textbf{2012} \textit{3}, 887, DOI: 10.1038/ncomms1882.
\bibitem{mak2012} K. F. Mak, K. He, J. Shan, T. F. Heinz, \textit{Nature nanotechnology}. \textbf{2012} \textit{7}, 494, DOI: 10.1038/nnano.2012.96.
\bibitem{yoo2015} S. Yoo, Q.-H. Park, \textit{Physical review letters}. \textbf{2015} \textit{114}, 203003, DOI: 10.1103/PhysRevLett.114.203003.
\bibitem{venkata2017} D. Venkatakrishnarao, C. Sahoo, E. A. Mamonov, V. B. Novikov, N. V. Mitetelo, S. R. G. Naraharisetty, T. V. Murzina, R. Chandrasekar, \textit{J. Mater. Chem. C.} \textbf{2017} \textit{5}, 12349, DOI: 10.1039/C7TC04621A.
\bibitem{stahl2020} P. Stahl, B. E. Arenas, S. R. Domingos, G. W. Fuchs, M. Schnell, T. F. Giesen, \textit{Physical chemistry chemical physics : PCCP.} \textbf{2020} \textit{22}, 21474, DOI: 10.1039/d0cp03523h.
\bibitem{gaul2020} K. Gaul, M. G. Kozlov, T. A. Isaev, R. Berger, \textit{Physical review letters}. \textbf{2020} \textit{125}, 123004, DOI: 10.1103/PhysRevLett.125.123004.
\bibitem{mak2018} K. F. Mak, Di Xiao, J. Shan, \textit{Nat. Photonics}. \textbf{2018} \textit{12}, 451, DOI: 10.1038/s41566-018-0204-6.
\bibitem{schaibley2016} J. R. Schaibley, H. Yu, G. Clark, P. Rivera, J. S. Ross, K. L. Seyler, W. Yao, X. Xu, \textit{Nat Rev Mater}. \textbf{2016} \textit{1}, DOI: 10.1038/natrevmats.2016.55.
\bibitem{collins2017} J. T. Collins, C. Kuppe, D. C. Hooper, C. Sibilia, M. Centini, V. K. Valev, \textit{Advanced Optical Materials}. \textbf{2017} \textit{5}, 1700182, DOI: 10.1002/adom.201700182.
\bibitem{wang2016a} Z. Wang, F. Cheng, T. Winsor, Y. Liu, \textit{Nanotechnology}. \textbf{2016} \textit{27}, 412001, DOI: 10.1088/0957-4484/27/41/412001.
\bibitem{passaseo2017} A. Passaseo, M. Esposito, M. Cuscunà, V. Tasco, \textit{Advanced Optical Materials}. \textbf{2017} \textit{5}, 1601079, DOI: 10.1002/adom.201601079.
\bibitem{hentschel2017} Hentschel Mario, Schäferling Martin, Duan Xiaoyang, Giessen Harald, Liu Na, \textit{Science Advances}. \textbf{2017} \textit{3}, e1602735, DOI: 10.1126/sciadv.1602735.
\bibitem{qiu2018} M. Qiu, L. Zhang, Z. Tang, W. Jin, C.-W. Qiu, D. Y. Lei, \textit{Adv. Funct. Mater.} \textbf{2018} \textit{28}, 1803147, DOI: 10.1002/adfm.201803147.
\bibitem{ayuso2019} D. Ayuso, O. Neufeld, A. F. Ordonez, P. Decleva, G. Lerner, O. Cohen, M. Ivanov, O. Smirnova, \textit{Nat. Photonics}. \textbf{2019} \textit{13}, 866, DOI: 10.1038/s41566-019-0531-2.
\bibitem{plum2015} E. Plum, N. I. Zheludev, \textit{Appl. Phys. Lett}. \textbf{2015} \textit{106}, 221901, DOI: 10.1063/1.4921969.
\bibitem{plum2016} E. Plum, \textit{Appl. Phys. Lett.} \textbf{2016} \textit{108}, 241905, DOI: 10.1063/1.4954033.
\bibitem{jing2017} L. Jing, Z. Wang, Y. Yang, B. Zheng, Y. Liu, H. Chen, \textit{Appl. Phys. Lett.} \textbf{2017} \textit{110}, 231103, DOI: 10.1063/1.4985132.
\bibitem{wang2016c} Z. Wang, H. Jia, K. Yao, W. Cai, H. Chen, Y. Liu, \textit{ACS Photonics}. \textbf{2016} \textit{3}, 2096, DOI: 10.1021/acsphotonics.6b00533.
\bibitem{ye2017} W. Ye, X. Yuan, C. Guo, J. Zhang, B. Yang, S. Zhang, \textit{Phys. Rev. Applied}. \textbf{2017} \textit{7}, DOI: 10.1103/PhysRevApplied.7.054003.
\bibitem{khurgin2015} J. B. Khurgin, \textit{Nature nanotechnology}. \textbf{2015} \textit{10}, 2, DOI: 10.1038/nnano.2014.310.
\bibitem{wu2014} C. Wu, N. Arju, G. Kelp, J. A. Fan, J. Dominguez, E. Gonzales, E. Tutuc, I. Brener, G. Shvets, \textit{Nature communications}. \textbf{2014} \textit{5}, 3892, DOI: 10.1038/ncomms4892.
\bibitem{khorasan2016} M. Khorasaninejad, A. Ambrosio, P. Kanhaiya, F. Capasso, \textit{Science Advances}. \textbf{2016} \textit{2}, e1501258, DOI: 10.1126/sciadv.1501258.
\bibitem{yin2015} X. Yin, M. Schäferling, A.-K. U. Michel, A. Tittl, M. Wuttig, T. Taubner, H. Giessen, \textit{Nano letters}. \textbf{2015} \textit{15}, 4255, DOI: 10.1021/nl5042325.
\bibitem{mai2019} W. Mai, D. Zhu, Z. Gong, X. Lin, Y. Chen, J. Hu, D. H. Werner, \textit{AIP Advances}. \textbf{2019} \textit{9}, 45305, DOI: 10.1063/1.5025560.
\bibitem{semnani2019} B. Semnani, J. Flannery, R. A. Maruf, M. Bajcsy, in: Proceedings 2019 Conference on Lasers and Electro-Optics Europe \& European Quantum Electronics Conference (CLEO/Europe-EQEC), Munich, Germany, 23.06.2019 - 27.06.2019, in \textit{2019 Conference on Lasers and Electro-Optics Europe \& European Quantum Electronics Conference (CLEO/Europe-EQEC)} (IEEE), p. 1, DOI: 10.1109/CLEOE-EQEC.2019.8872856.
\bibitem{zhu2018} A. Y. Zhu, W. T. Chen, A. Zaidi, Y.-W. Huang, M. Khorasaninejad, V. Sanjeev, C.-W. Qiu, F. Capasso, \textit{Light, science \& applications}. \textbf{2018} \textit{7}, 17158, DOI: 10.1038/lsa.2017.158.
\bibitem{liu2012} V. Liu, S. Fan, \textit{Computer Physics Communications}. \textbf{2012} \textit{183}, 2233, DOI: 10.1016/j.cpc.2012.04.026.
\bibitem{li1997} L. Li, \textit{J. Opt. Soc. Am. A}. \textbf{1997} \textit{14}, 2758, DOI: 10.1364/JOSAA.14.002758.
\bibitem{wang2016b} Z. Wang, H. Jia, K. Yao, W. Cai, H. Chen, Y. Liu, \textit{ACS Photonics}. \textbf{2016} \textit{3}, 2096, DOI: 10.1021/acsphotonics.6b00533.
\bibitem{aspnes1983} D. E. Aspnes, A. A. Studna, \textit{Phys. Rev. B}. \textbf{1983} \textit{27}, 985, DOI: 10.1103/PhysRevB.27.985.
\bibitem{hansen2001} N. Hansen, A. Ostermeier, \textit{Evolutionary computation}. \textbf{2001} \textit{9}, 159, DOI: 10.1162/106365601750190398.
\bibitem{hansen2003} N. Hansen, S. D. Müller, P. Koumoutsakos, \textit{Evolutionary computation}. \textbf{2003} \textit{11}, 1, DOI: 10.1162/106365603321828970.
\bibitem{shapley2016} L. S. Shapley, in \textit{Contributions to the Theory of Games (AM-28), Volume II} Princeton University Press \textbf{2016}, pp. 307–318, DOI: 10.1515/9781400881970-018.
\bibitem{lundberg2017} S. M. Lundberg, S.-I. Lee, in:, in \textit{Proceedings of the 31st International Conference on Neural Information Processing Systems}, NIPS’17 (Curran Associates Inc, Red Hook, NY, USA, 2017), pp. 4768–4777.
\bibitem{erhan2009} D. Erhan, Y. Bengio, A. Courville, P. Vincent, \textit{Visualizing Higher-Layer Features of a Deep Network}, \textbf{2009}.
\bibitem{zeiler2014} M. D. Zeiler, R. Fergus, in: \textit{Computer Vision ‐ ECCV 2014} (Springer International Publishing, Cham, 2014), pp. 818–833, DOI: 10.1007/978-3-319-10590-1\_53.
\bibitem{oskooi2010} A. F. Oskooi, D. Roundy, M. Ibanescu, P. Bermel, J. D. Joannopoulos, S. G. Johnson, \textit{Computer Physics Communications}. \textbf{2010} \textit{181}, 687, DOI: 10.1016/j.cpc.2009.11.008.
\end{thebibliography}
\end{document}